%% file: main.tex
\documentclass[conference]{IEEEtran}
\IEEEoverridecommandlockouts
\usepackage{cite}
\usepackage{amsmath,amssymb,amsfonts}
\usepackage{algorithmic}
\usepackage{graphicx}
\usepackage{textcomp}
\usepackage{xcolor}
\usepackage{pifont}
\usepackage[T1]{fontenc}
\usepackage{booktabs}  
\usepackage{multirow}
\usepackage{stfloats}
\usepackage{tcolorbox}
\tcbuselibrary{skins}
\usepackage{framed}

\def\BibTeX{{\rm B\kern-.05em{\sc i\kern-.025em b}\kern-.08em
    T\kern-.1667em\lower.7ex\hbox{E}\kern-.125emX}}

\begin{document}
\title{Are GNNs Actually Effective for Multimodal Fault Diagnosis in Microservice Systems?}
\author{
\IEEEauthorblockN{Fei Gao\textsuperscript{1}, Ruyue Xin\textsuperscript{2}, Xiaocui Li\textsuperscript{2,3}, and Yaqiang Zhang\textsuperscript{1}}
\IEEEauthorblockA{
\textsuperscript{1}\textit{Inspur (Beijing) Electronic Information Industry Co., Ltd.}, Beijing, China \\
\textsuperscript{2}\textit{China University of Geosciences}, Beijing, China \\
\textsuperscript{3}\textit{Peking University}, Beijing, China \\
\{gaofei06, zhangyaqiang\}@ieisystem.com, rxin@cugb.edu.cn, lixiaocui@pku.edu.cn
}}
\maketitle

\begin{abstract}
Graph Neural Networks (GNNs) are widely adopted for fault diagnosis in microservice systems, premised on their ability to model service dependencies. However, the necessity of explicit graph structures remains underexamined, as existing evaluations conflate preprocessing with architectural contributions. To isolate the true value of GNNs, we propose DiagMLP—a deliberately minimal, topology-agnostic baseline that retains multimodal fusion capabilities while excluding graph modeling. Through ablation experiments across five datasets, DiagMLP achieves performance parity with state-of-the-art GNN-based methods in fault detection, localization, and classification. These findings challenge the prevailing assumption that graph structures are indispensable, revealing that: (i) preprocessing pipelines already encode critical dependency information, and (ii) GNN modules contribute marginally beyond multimodality fusion. Our work advocates for systematic re-evaluation of architectural complexity and highlights the need for standardized baseline protocols to validate model innovations.
\end{abstract}


\section{Introduction}

Microservice systems (MicroSS) have become the cornerstone of modern cloud-based distributed applications, offering unparalleled scalability, flexibility, and modularity. These systems are composed of numerous interdependent services that generate multimodal telemetry data (e.g., logs, metrics, traces) for continuous monitoring. Diagnosing faults in MicroSS is particularly challenging due to two inherent complexities: cross-modal symptom correlations where faults manifest through multiple data modalities simultaneously, and dynamic fault propagation along service dependency graphs \cite{Zhang2024Survey, Xin2023}.

To address these challenges, recent studies \cite{Eadro,DiagFusion,TVDiag,Zhang2024MASS,DeepHunt} have explored Graph Neural Networks (GNNs) for jointly modeling telemetry data with service topology. Despite their prevalence, current empirical evaluations lack rigorous ablation studies to isolate the contributions of preprocessing, embedding, and modeling stages. For example, trace-based preprocessing may already encode dependency graphs \cite{DiagFusion}, potentially overshadowing GNN-specific benefits. This raises concerns that performance gains in GNN-based works may stem from advanced preprocessing and embedding pipeline rather than architectural innovation.

\begin{figure}[tbp] 
\centerline{\includegraphics[width=\linewidth]{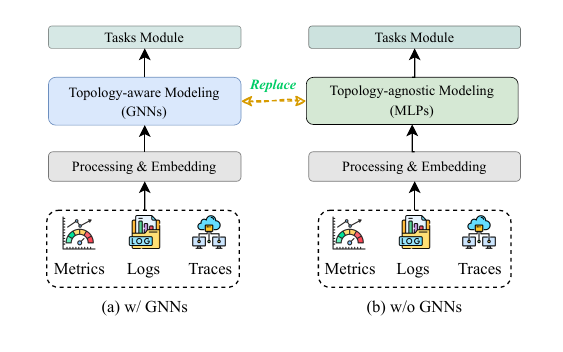}} \caption{\textbf{Overview of GNNs ablation methods.} (a) shows fault diagnosis using GNNs as the backbone to model the service dependency graphs. (b) shows the model with GNNs replaced by a topology-agnostic MLPs while keeping the other components.}
\label{background_fig} 
\end{figure}

To rigorously assess these issues, we introduce \textbf{DiagMLP, a ablation experimental baseline rather than a competing model}. By maintaining identical preprocessing and embedding pipelines as GNN-based methods, DiagMLP preserves the multimodal fusion capabilities inherent to existing frameworks while \textbf{explicitly removing service dependency graph modeling}.  Furthermore, DiagMLP retains the same task modules (detection, localization, classification) as prior works, enabling fair comparisons under consistent experimental conditions. The overview of our ablation experiment design is illustrated in Fig. \ref{background_fig}.

Our experiments reveal that DiagMLP matches GNN-based performance across fault detection, localization, and classification tasks, challenging the necessity of explicit graph modeling. This parity suggests preprocessing pipelines already capture critical dependency information, diminishing the marginal utility of GNNs. These findings advocate for three paradigm shifts: (i) rigorous baseline comparisons to validate architectural innovations, (ii) standardized protocols to isolate preprocessing effects, and (iii) datasets with explicit dependency-driven fault patterns to benchmark future models.

The paper is organized as follows: Section II reviews related work and identifies their limitations, Section III introduces DiagMLP, our topology-agnostic baseline, Section IV presents experiments and ablation studies, Section V discusses results, and Section VI concludes.

\section{GNN-Based Fault Diagnosis Methods}

\begin{figure*}[htp!]
\centerline{\includegraphics[width=0.8\textwidth]{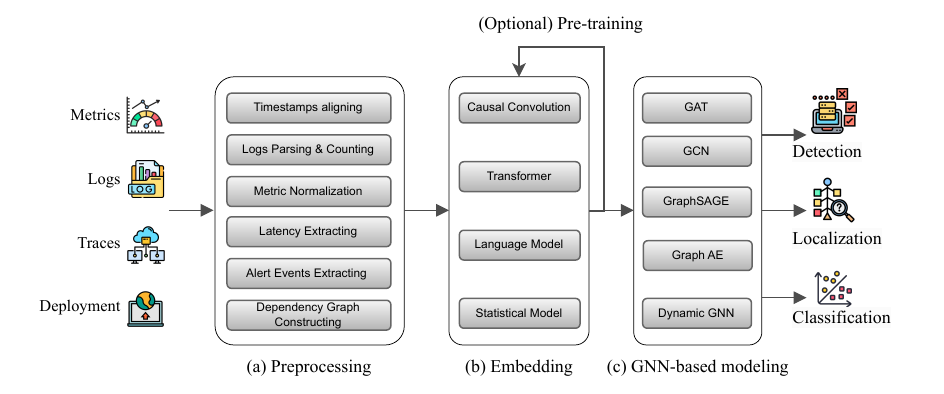}}
\caption{\textbf{Pipeline of Existing GNN-Based Multimodal Fault Diagnosis Models.} (a) Preprocessing transforms raw multimodal data (metrics, logs, traces) into standardized formats, such as time series, text, and graphs; (b) Embedding modules, often pretrained, encode these features into dense representations; (c) GNN modules, varying across methods, integrate multimodal features with dependency graphs for diagnosis.}
\label{fig:current_research}
\end{figure*}

Existing methods for fault diagnosis in MicroSS typically construct dependency graphs from trace data and deployment configurations, fuse multimodal telemetry (metrics, logs, traces) into node features, and apply GNNs for diagnosis \cite{DeepTraLog, MSTGAD}. Fig. \ref{fig:current_research} outlines this pipeline, comprising preprocessing, embedding, and GNN-based modeling.

\paragraph{\textbf{Preprocessing}}
Raw multimodal data—metrics, logs, and traces—requires preprocessing due to its unstandardized nature. Logs are parsed into templates using tools like Drain3 \cite{Drain} and transformed into time series of template occurrences through heuristic filtering \cite{TVDiag, MSTGAD, CHASE}, reducing redundancy while preserving key patterns. Metrics, as multivariate time series, are standardized and often compressed via clustering \cite{Eadro} for efficiency. Traces provide latency features and, when paired with deployment configurations, enable construction of service dependency graphs \cite{DiagFusion, DeepHunt}. To address anomaly sparsity, lightweight techniques like the 3-sigma rule \cite{TVDiag} generate alert event sequences, offering a unified, informative representation across modalities \cite{DeepTraLog, Nezha}.

\paragraph{\textbf{Embedding}}
Embedding preprocessed data captures temporal dynamics and semantic information. Time-series data from metrics and logs is encoded using sequence models such as Temporal Convolutional Networks (TCNs) \cite{TCN} or Transformers \cite{Transformer} to model complex dependencies. Alert sequences are embedded with language models like FastText \cite{FastText} or GloVe \cite{GloVe}, while raw logs may leverage pretrained BERT \cite{BERT} for contextual richness. Embedding strategies vary, with some methods pretraining modules \cite{DiagFusion, TVDiag} and others integrating them end-to-end in the pipeline.

\paragraph{\textbf{GNN-based Modeling}} 
GNNs excel at modeling fault propagation along service dependencies, a key challenge in MicroSS diagnosis. Standard architectures like Graph Convolutional Networks (GCNs), Graph Attention Networks (GATs) \cite{GAT}, and GraphSAGE are widely used, while advanced variants address specific needs: DGERCL \cite{DGERCL} employs dynamic GNNs for temporal service invocation sequences, and DeepHunt \cite{DeepHunt} introduces a graph autoencoder for self-supervised learning amid limited labeled data. These approaches integrate multimodal features with graph topology to enhance diagnostic accuracy.

\paragraph{\textbf{Observations and Motivation}}
Two observations emerge from this review. First, while preprocessing and embedding techniques differ significantly, GNNs dominate the modeling stage, yet their added value lacks rigorous benchmarking against simpler baselines. Second, preprocessing—especially trace processing—often embeds dependency information into features, potentially overshadowing the unique contributions of GNNs. These findings prompt a critical question: does the complexity of GNNs justify their widespread adoption in MicroSS fault diagnosis, given the advanced preprocessing and embedding pipelines?

\section{A Topology-Agnostic Baseline}
State-of-the-art GNN-based models typically optimize the entire fault diagnosis pipeline, obscuring whether performance gains stem from GNN modules or from other components, such as multimodal data fusion and preprocessing. \textbf{We hypothesize that non-GNN elements primarily drive these improvements. }To test this, we propose \textit{DiagMLP}, a topology-agnostic baseline designed to isolate and evaluate the true contributions of GNNs by excluding explicit graph modeling.

\subsection{Problem Formulation}
Given a MicroSS represented as a set of $N$ nodes (service instances), each node $i \in \{1, \dots, N\}$ is associated with multimodal features:  
\begin{equation}
    \mathbf{x}_i = \left[ \mathbf{x}_i^{\text{metric}}, \mathbf{x}_i^{\text{log}}, \mathbf{x}_i^{\text{trace}} \right],
\end{equation}
where $\mathbf{x}_i^{\text{metric}}, \mathbf{x}_i^{\text{log}}, \mathbf{x}_i^{\text{trace}} \in \mathbb{R}^d $ are embeddings derived from metric, log, and trace data, respectively. These embeddings can be generated using any preprocessing and embedding pipeline, without imposing restrictions on the specific methods employed.

The goal is to learn a mapping $ f: \{\mathbf{x}_i\}_{i=1}^N \rightarrow \mathbb{R}^{c} $ that determines whether the system is anomalous (for $c=2$), identifies the root cause (for $c=N$), and classifies the type of failure (where $c$ is the number of failure types).

\subsection{A Motivating Example}
\begin{figure}
    \centering
    \includegraphics[width=1\linewidth]{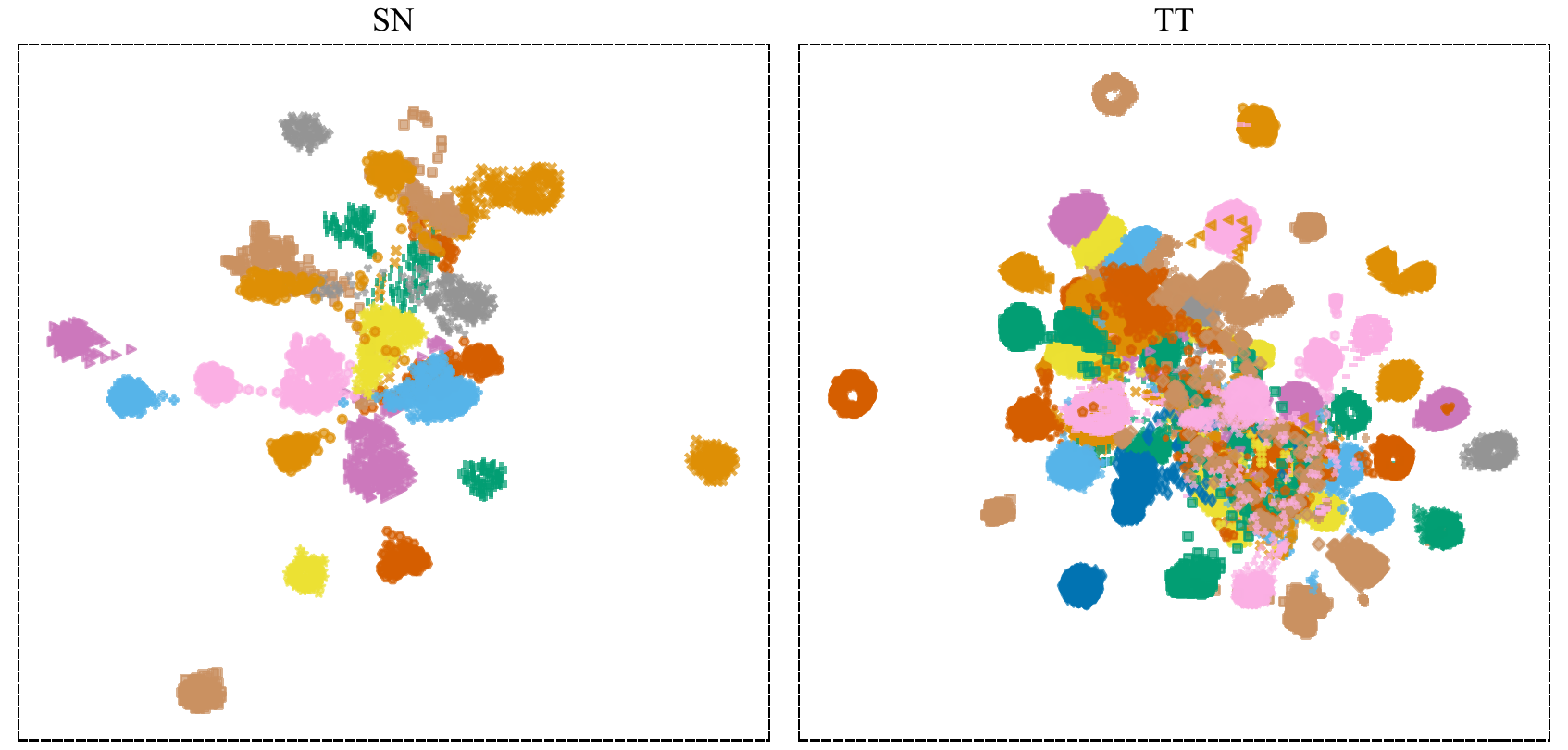}
    \caption{\textbf{Motivating Example.} UMAP visualization of preprocessed multimodal features from the SN and TT datasets \cite{Eadro}. Each point represents a fault window, colored by its root cause, showing clear clustering without topology modeling.}
    \label{fig:sntt}
\end{figure}

To justify our ablation design, we analyze the Social Networks (SN) and Train Ticket (TT) datasets from Eadro \cite{Eadro}, focusing on the inherent separability of preprocessed multimodal features. For each time window $t$, we concatenate features from all nodes into a system-level representation: $\mathbf{X}_t = [\mathbf{x}_1, \mathbf{x}_2, \dots, \mathbf{x}_N ]\in \mathbb{R}^{3dN}$. This topology-agnostic feature $\mathbf{X}_t$ excludes explicit service dependency modeling. Each fault window $\mathbf{X}_t$ is labeled with its root cause $r_t \in \{1, 2, \dots, N\}$. Using UMAP \cite{UMAP}, we project $\mathbf{X}_t$ from from $\mathbb{R}^{3dN}$ to $\mathbb{R}^{2}$ and visualize the results in Fig. \ref{fig:sntt}, coloring points by $r_t$.

The visualization reveals that fault windows with the same root cause cluster tightly and separate distinctly from others. This suggests that topology-agnostic features $\mathbf{X}_t$ already encode sufficient information for root cause identification in the SN and TT datasets, raising doubts about the necessity of GNN-based topology modeling and motivating our baseline approach.

\subsection{Topology-agnostic DiagMLP}
\begin{figure}[tbp!]
\label{fig:related_works}
\centerline{\includegraphics[width=1\linewidth]{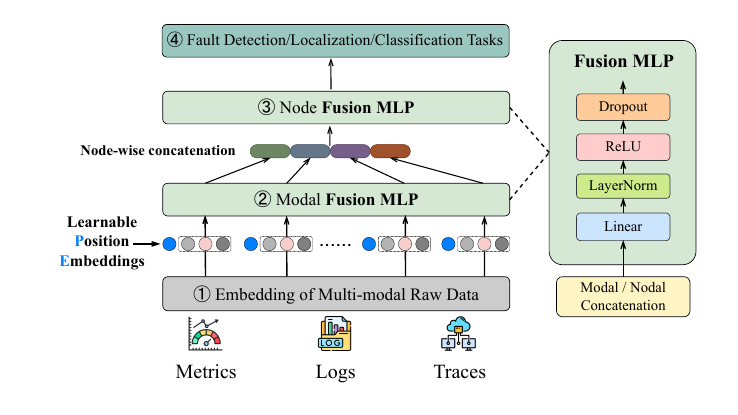}}
\caption{\textbf{DiagMLP Architecture.} DiagMLP (\ding{193}+\ding{194}) replaces GNN modules, positioned between any embedding module (\ding{192}) and downstream task module (\ding{195}), processing multimodal features without topology modeling.}
\label{fig:diagmlp}
\end{figure}

We introduce \textit{DiagMLP}, a minimal fault diagnosis model based on multi-layer perceptrons (MLPs), designed as a baseline to evaluate the necessity of GNNs in MicroSS. DiagMLP is topology-agnostic, meaning it does not model the dependency graph of service interactions (i.e., the topological structure typically used by GNNs to capture relationships between services). Instead, it processes node features independently, focusing solely on multimodal fusion. Its architecture, shown in Fig. \ref{fig:diagmlp}, is defined as:
\begin{equation}
\label{eq:diagmlp}
\begin{aligned}
    \mathbf{x}_i &= \mathbf{x}_i^{\text{metric}} \oplus \mathbf{x}_i^{\text{log}} \oplus \mathbf{x}_i^{\text{trace}} \oplus p_i, \\
    \mathbf{x}_i &= \texttt{Modal\_Fusion\_MLP}(\mathbf{x}_i), \\
    x_g &= x_1 \oplus x_2 \oplus \cdots \oplus x_N, \\
    y_g &= \texttt{Node\_Fusion\_MLP}(x_g),
\end{aligned}
\end{equation}
where $\oplus$ denotes the concatenation operator, $p_i \in \mathbb{R}^d $ is the learnable position embedding \cite{PE} for node $i$, and $y_g$ is the system-level representation. Both \texttt{Fusion\_MLP} modules share a simple structure:
\begin{equation}
\label{equ:mlp}
\texttt{Fusion\_MLP}(x) = \texttt{Dropout} \big( \texttt{ReLU} \big( \texttt{LN}(Wx + b) \big) \big),
\end{equation}
with layer normalization (\texttt{LN}), \texttt{ReLU} activation, and dropout for regularization.

By avoiding the explicit modeling of service dependencies, DiagMLP (\ding{193} and \ding{194}) maps node-level embeddings to system-level outputs, serving as a flexible replacement for GNN modules in existing pipelines. It imposes no constraints on the embedding module (\ding{192}) or downstream tasks (\ding{195}), ensuring compatibility with diverse frameworks.

\subsection{Design Choices of DiagMLP}

DiagMLP design balances simplicity and efficacy through deliberate choices:

\paragraph{Learnable Position Embeddings}
Each node is assigned a learnable position embedding $p_i$, enhancing performance by improving node distinguishability. This likely aids in capturing node-specific fault patterns, compensating for the absence of topological structure.

\paragraph{Single-Layer MLP}
Both fusion MLPs use a single-layer architecture. This minimizes complexity, reduces overfitting risks, and highlights the contributions of preprocessing and embedding stages over intricate modeling.

\paragraph{Fusion by Concatenation}
Modal-wise and node-wise features are integrated through straightforward concatenation to retain the full breadth of input information. While this method proves effective, it results in the parameter space of the \texttt{Node\_Fusion\_MLP} scaling linearly with the number of nodes $N$. Although this complexity is manageable for publicly available multimodal datasets, it poses significant challenges to scalability in larger systems. To overcome these limitations, future research could investigate more sophisticated and scalable alternatives, such as DeepSets \cite{deepset} or Set Transformers \cite{Settransformer}.

\input{tables/data_table.tex}

\section{Experiments}
To test our hypothesis, that performance gains in SOTA GNN-based fault diagnosis frameworks stem primarily from non-GNN components, we replace GNN modules with DiagMLP while preserving all other pipeline elements, including raw datasets, preprocessing, embedding techniques, objectives, and training strategies. This ablation design isolates the modeling stage, ensuring that performance differences reflect the contribution of GNNs alone, free from confounding factors.

\subsection{Experimental Settings}
\paragraph{\textbf{Baselines}}
We integrate DiagMLP into three recent SOTA frameworks—Eadro \cite{Eadro}, TVDiag \cite{TVDiag}, and DeepHunt \cite{DeepHunt},by substituting their GNN components with our topology-agnostic model. This allows us to assess the effectiveness of DiagMLP within established pipelines. For broader comparison, we also include DiagFusion \cite{DiagFusion} and CHASE \cite{CHASE}, which share compatible evaluation protocols.

\paragraph{\textbf{Datasets}}  We adopt datasets from the original studies to ensure consistency: Social Network (\textbf{SN}) and Train Tickets (\textbf{TT}) from Eadro\footnote{https://zenodo.org/doi/10.5281/zenodo.7615393}  , \textbf{GAIA} from TVDiag\footnote{https://github.com/WHU-AISE/TVDiag}, and \textbf{D1} and \textbf{D2} from DeepHunt\footnote{https://github.com/bbyldebb/DeepHunt}. For SN and TT, we reimplemented preprocessing code due to missing original scripts, while for GAIA, D1, and D2, we used the preprocessed multimodal data from the authors directly. Dataset statistics are summarized in Table \ref{tab:dataset}.

\paragraph{\textbf{Evaluation Metrics}} For fault detection and classification, we report precision, recall, and F1-score. For fault localization, we use Top-$k$ accuracy, measuring the fraction of cases where the faulty node appears in the top $k$ predictions.

\paragraph{\textbf{Notes}} (1) We limit our evaluation to the datasets and models relevant to our ablation study, avoiding exhaustive cross-testing irrelevant to isolating GNN effects. (2) Comparisons with machine learning or non-GNN deep learning approaches are omitted, as they diverge from our objective of assessing GNN-specific contributions. (3) We rectified issues in prior frameworks, such as the  window-splitting method in Eadro \cite{Eadro}, which risked data leakage, and incorporated a validation set to meet standard machine learning evaluation criteria. (4) Our experiments leverage the most comprehensive open-source multimodal fault diagnosis datasets available; larger, more complex datasets were excluded due to their proprietary nature, not methodological limitations.

\subsection{Results}

\input{tables/fd_table.tex}
\input{tables/fc_table.tex}
\input{tables/fl_res.tex}

We present the fault detection results for the SN and TT datasets in Table \ref{tab:fd}, fault classification results for the GAIA dataset in Table \ref{tab:fc}, and fault localization results across all datasets (SN, TT, GAIA, D1, D2) in Table \ref{tab:fl}.
DiagMLP consistently achieves competitive or superior performance across all tasks compared to SOTA GNN-based models, supporting our hypothesis that GNN modules contribute minimally to overall performance. In fault detection (Table \ref{tab:fd}), DiagMLP outperforms Eadro on both SN (F1: 96.7\% vs. 92.1\%) and TT (F1: 90.8\% vs. 90.7\%), with notable gains in recall (e.g., 99.7\% vs. 92.0\% on SN). This suggests that our topology-agnostic approach captures critical fault patterns effectively, without relying on graph structures.

For fault classification on GAIA (Table \ref{tab:fc}), DiagMLP (F1: 93.4 ± 0.2\%) performs comparably to TVDiag (F1: 94.5 ± 0.1\%), with differences within standard deviations, indicating no significant advantage from  GraphSAGE \cite{sage} module in TVDiag. Similarly, in fault localization (Table \ref{tab:fl}), DiagMLP surpasses its backbones across all datasets. On SN and TT with Eadro as the backbone, DiagMLP achieves substantial improvements (e.g., SN Top-1: 80.2 ± 3.1\% vs. 41.8 ± 14.5\%; TT Top-1: 98.5 ± 0.8\% vs. 91.1 ± 2.2\%). The high variance of Eadro (e.g., ±14.5\% on SN Top-1) suggests overfitting, likely due to its complex GAT \cite{GAT} architecture, whereas the simplicity of DiagMLP enhances robustness. On GAIA, D1, and D2, with TVDiag and DeepHunt backbones, DiagMLP matches or exceeds performance (e.g., D1 Top-1: 79.8 ± 0.4\% vs. DeepHunt’s 31.0\%), further highlighting the limited marginal utility of GNNs.

\textit{\textbf{Key Insight 1:}} The superior or comparable performance of DiagMLP highlights the limited benefits of GNNs in fault diagnosis, a gap overlooked by prior work lacking rigorous GNNs ablation studies.

\subsection{Visualization}

\begin{figure}
    \centering
    \includegraphics[width=\linewidth]{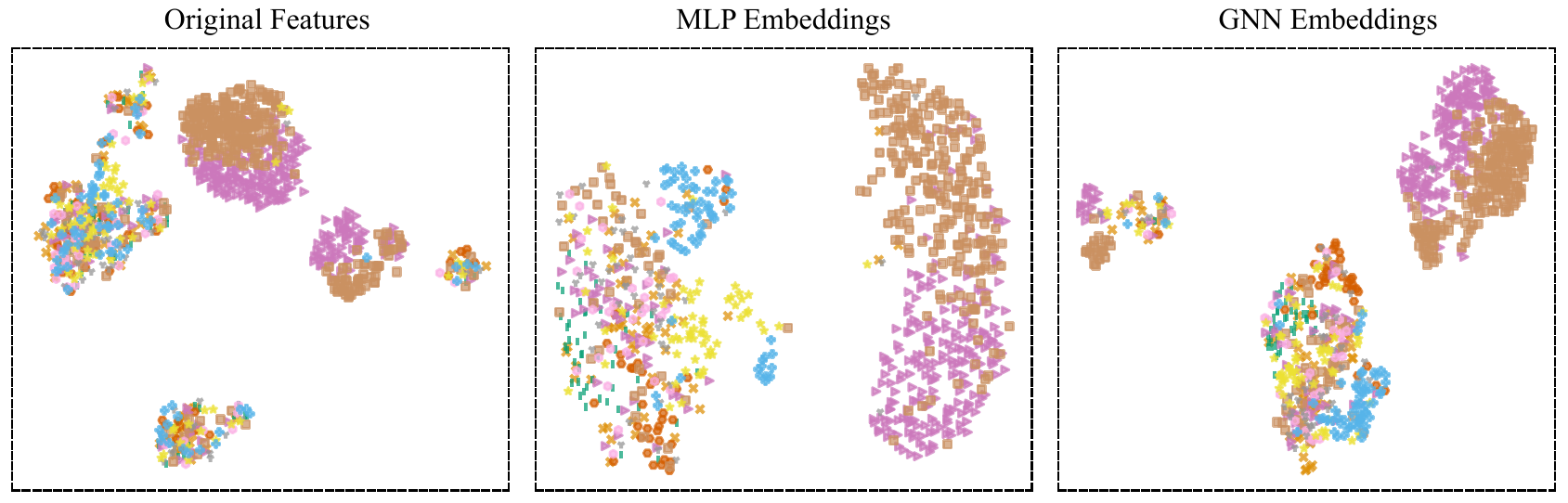}
    \caption{\textbf{UMAP Visualization of GAIA Dataset.} Fault windows are depicted using node-wise \textit{Original Features} (from FastText), \textit{MLP Embeddings} (from DiagMLP), and \textit{GNN Embeddings} (from TVDiag).}
    \label{fig:gaia}
\end{figure}
We employ UMAP \cite{UMAP} to visualize embeddings across datasets, assessing their ability to distinguish fault characteristics. Building on the earlier motivating example (Fig. \ref{fig:sntt}), where concatenated preprocessed multimodal features from SN and TT datasets effectively separated fault root causes, we extend this analysis to the GAIA dataset. In Fig. \ref{fig:gaia}, we compare three representations: (1) \textit{Original Features} from the embedding module (FastText), (2) \textit{MLP Embeddings} from the node fusion outputs in DiagMLP, and (3) \textit{GNN Embeddings} from the graph pooling outputs in TVDiag. The visualizations reveal that neither DiagMLP nor TVDiag significantly enhances class separation beyond the original features. This aligns with SN and TT findings, where topology-agnostic features alone sufficed, suggesting that topological modeling via GNNs provides minimal additional discriminative power.

\textit{\textbf{Key Insight 2:}} The richness of preprocessed features enables DiagMLP’s effectiveness. Both MLP and GNN embeddings fail to enhance fault separability, explaining the limited effectiveness of GNNs.

\section{Analysis and Discussion}

The competitive performance of the topology-agnostic DiagMLP, as evidenced by its superior results, suggests that current SOTA methods fail to fully exploit the assumed benefits of GNN-based dependency modeling. Moreover, the high variance of Eadro indicates that GNN complexity may even degrade accuracy. We propose two reasons: (1) topological dependencies remain underutilized or redundant, as prior studies \cite{Luo2021, Zhang2024Survey} suggest potential in related domains yet show limited impact here; (2) preprocessing embeds essential topological information, rendering explicit GNN modeling unnecessary, supported by UMAP visualizations showing minimal separability improvement. The small scale of datasets like SN and TT (e.g., dozens of instances) may further limit the relevance of global topology.

These findings urge the MicroSS research community to cautiously adopt complex GNN models and prioritize rigorous ablation studies, a gap in existing work. We recommend developing larger datasets (e.g., with 1000+ service instances) and standardized preprocessing protocols (e.g., time alignment, feature normalization) to better evaluate model innovations. Adopting robust validation practices will enhance reproducibility and generalizability.

\section{Conclusion}
This study conducts an ablation analysis to assess the necessity of GNNs in microservice system fault diagnosis, using a minimal topology-agnostic baseline, DiagMLP. Experiments across five datasets show DiagMLP achieving parity with state-of-the-art GNN-based methods, while visualizations reveal limited GNNs contribution beyond preprocessing. Findings indicate that multimodal fusion and preprocessing, not graph structures, mainly drive the performance, challenging GNNs reliance. The absence of standardized baselines and complex datasets highlights the necessity for a rigorous reassessment of architectural complexity within future fault diagnosis research.

\end{document}

%% file: tables/data_table.tex
\begin{table}[htb]
    \centering
    \caption{Statistics of the datasets}
    \label{tab:dataset}
    \begin{minipage}{\linewidth} 
        \centering
        \renewcommand{\thefootnote}{\alph{footnote}} 
        \begin{tabular}{c|c|ccc|c}
        \toprule
            Dataset & \#instances & \#train & \#valid & \#test & Task \footnote{Fault Detection (Det.), Localization (Loc.) and Classification (Cla.).}  \\
        \midrule
            \textbf{SN}   & 12 & 316   & 78  & 169  & Det. \& Loc.\\
            \textbf{TT}   & 27 & 3256  & 813 & 1744 & Det. \& Loc. \\
            \textbf{GAIA} & 10 & 128   & 32  & 939  & Cla. \& Loc. \\
            \textbf{D1}   & 46 & 63    & -   & 147  & Loc. \\
            \textbf{D2}   & 18 & 40    & -   & 93   & Loc. \\
        \bottomrule
        \end{tabular}
        \vspace{0em} 
    \end{minipage}
\end{table}

%% file: tables/fd_table.tex
\begin{table}[t]
\centering
\caption{Fault Detection results for SN and TT dataset}
\label{tab:fd}
\begin{tabular}{c|ccc|ccc}
\toprule
& \multicolumn{3}{c|}{\textbf{SN}}       & \multicolumn{3}{c}{\textbf{TT}}  \\
& Pre                    & Rec                       & F1                            & Pre                & Rec                       & F1                       \\
\midrule
\textbf{Eadro}\cite{Eadro}                       & 92.1                         & 92.0                         & 92.1                          & 83.3                     & 99.5                         & 90.7                     \\
\multicolumn{1}{l|}{\textbf{DiagMLP}} & \multicolumn{1}{l}{93.8} & \multicolumn{1}{l}{99.7} & \multicolumn{1}{l|}{96.7} & \multicolumn{1}{l}{83.4} & \multicolumn{1}{l}{99.8} & \multicolumn{1}{l}{90.8} \\
\bottomrule
\end{tabular}
\end{table}

%% file: tables/fc_table.tex
\begin{table}[ht] 
\centering
\caption{Fault classification results for GAIA dataset}
\label{tab:fc}
\begin{tabular}{c|ccc}
\toprule
Methods & Pre & Rec & F1 \\
\midrule
\textbf{TVDiag} \cite{TVDiag} & 93.4 ± 0.2  & 95.5 ± 0.0 & 94.5 ± 0.1  \\
\textbf{DiagMLP}  & 92.1 ± 0.2  & 94.7 ± 0.3 & 93.4 ± 0.2 \\
\bottomrule
\end{tabular}
\end{table}

%% file: tables/fl_res.tex
\begin{table*}[htb]
\centering
\begin {minipage} {0.9\textwidth}
\centering
    \caption{Accuracy of Fault Localization in Microservice Systems }
    \label{tab:fl}
    \begin{tabular}{c|c|ccccc|cc}
    \toprule
                  \textbf{Datasets \footnote{Results without standard deviations are directly extracted from the cited papers, while results with standard deviations are derived from 10 independent experiments conducted under the same settings as the cited works.}}           &      & \textbf{Eadro} \cite{Eadro}         & \textbf{DiagFusion} \cite{DiagFusion}  & \textbf{TVDiag} \cite{TVDiag}     & \textbf{CHASE} \cite{CHASE}  & \textbf{DeepHunt} \cite{DeepHunt}     & \textbf{DiagMLP} (Ours) & \textbf{Backbone \footnote{Backbone refers to the original architecture of the cited method (Eadro, TVDiag, DeepHunt), with its GNN module substituted by DiagMLP to isolate the contribution of graph-based modeling.}} \\
    \midrule
    \multirow{3}{*}{\textbf{SN}}    & Top1 & 41.8 ± 14.5 &      -      &  -          &   -    &       -       & \textbf{80.2 ± 3.1}  & \multirow{3}{*}{Eadro}\\
                                 & Top3 & 61.1 ± 16.0 &      -      &   -         &    -   &       -       & \textbf{88.6 ± 2.8}  \\
                                 & Top5 & 71.7 ± 10.8 &       -     &  -          &   -    &      -        & \textbf{89.4 ± 3.2}  \\
    \midrule
    \multirow{3}{*}{\textbf{TT}}    & Top1 & 91.1 ± 2.2  &      -      &   -         &    -   &       -       & \textbf{98.5 ± 0.8}  & \multirow{3}{*}{Eadro}\\
                                 & Top3 & 96.5 ± 0.7  &       -     &    -        &   -    &         -     & \textbf{99.7 ± 0.6}  \\
                                 & Top5 & 97.7 ± 0.6  &      -      &     -       &   -    &        -      & \textbf{99.2 ± 0.6}  \\
    \midrule
    \multirow{3}{*}{\textbf{GAIA}}        & Top1 &       30.4        &     41.9       & 59.8 ± 0.8 &   \textbf{61.4}    &      -        & \textbf{61.0 ± 1.0}  & \multirow{3}{*}{TVDiag}\\
                                 & Top3 &       59.2       &    81.3        & 80.4 ± 0.5 &    \textbf{88.2}   &        -      & 80.7 ± 0.8  \\
                                 & Top5 &      -         &     \textbf{91.4}       & 88.5 ± 0.2 &   -    &        -      & 87.4 ± 0.8  \\
    \midrule
    \multirow{3}{*}{\textbf{D1}} & Top1 & 31.0          & 33.3       &  -          &    -   & \textbf{79.8 ± 0.4} & \textbf{79.8 ± 1.0} & \multirow{3}{*}{DeepHunt}\\
                                 & Top3 & 44.6          & 50.5       &  -          &    -   & \textbf{90.6 ± 0.6} & \textbf{90.6 ± 0.5 }  \\
                                 & Top5 & 48.4          & 64.8       &    -        &   -    & \textbf{96.7 ± 0.3} & \textbf{95.8 ± 1.0}  \\
    \midrule
    \multirow{3}{*}{\textbf{D2}} & Top1 & 21.4          & 39.8       &   -         &   -    & \textbf{78.9 ± 0.7} & \textbf{79.0 ± 0.5}  & \multirow{3}{*}{DeepHunt}\\
                                 & Top3 & 38.6          & 55.2       &   -         &   -    & \textbf{93.4 ± 0.3} & \textbf{93.0 ± 0.7}  \\
                                 & Top5 & 45.4          & 75.0       &   -         &    -   & \textbf{94.6 ± 0.0} & \textbf{94.6 ± 0.0} \\
    \bottomrule
    \end{tabular}
\end{minipage}
\end{table*}